\documentclass{aa}  

\usepackage{graphicx}
\usepackage{orcidlink}
\usepackage{xcolor}
\usepackage[title]{appendix}
\usepackage{ulem}
\usepackage{txfonts}
\begin{document}

   \title{First Solar Orbiter observation of a dark halo in the solar atmosphere}

   \author{S. M. Lezzi\,\orcidlink{0000-0002-8506-1819}\inst{1}\fnmsep\inst{2}, D. M.  Long\,\orcidlink{0000-0003-3137-0277}\inst{3},  V. Andretta\,\orcidlink{0000-0003-1962-9741}\inst{2}, D. Baker\,\orcidlink{0000-0002-0665-2355}\inst{4}, A. Dolliou\,\orcidlink{0000-0001-9379-9699} \inst{5}, M. Murabito\,\orcidlink{0000-0002-0144-2252}\inst{6,7}, S. Parenti\,\orcidlink{0000-0003-1438-1310}\inst{5} \and N. Zambrana Prado\,\orcidlink{0000-0001-6395-7115}\inst{4}
          }

      \institute{Università di Napoli “Federico II”, C.U. Monte Sant’Angelo, via Cinthia, I-80126 Napoli, Italy, \texttt{serena.lezzi@inaf.it}
         \and
             INAF Istituto Nazionale di Astrofisica, Osservatorio Astronomico di Capodimonte, Salita Moiariello 16, I-80131 Napoli, Italy
        \and 
             School of Physical Sciences, Dublin City University, Glasnevin Campus, Dublin D09V209, Ireland
        \and 
             University College London, Mullard Space Science Laboratory, Holmbury St. Mary, Dorking, Surrey, RH5 6NT, UK
        \and 
             Institut d’Astrophysique Spatiale, Bâtiment 121, Rue Jean Dominique Cassini, Université Paris Saclay, 91405 Orsay, France
        \and 
             INAF Istituto Nazionale di Astrofisica, Osservatorio Astronomico di Roma, I-00078 Monte Porzio Catone, Roma, Italy
        \and 
            Space Science Data Center, Agenzia Spaziale Italiana, via del Politecnico, s.n.c., I-00133, Roma, Italy
             }

   \date{Received June 17, 2024; accepted August 29, 2024}

 
\abstract
   {Solar active regions (ARs) are often surrounded by dark large areas of reduced emission compared to the quiet Sun, observed at various wavelengths corresponding to chromosphere, transition region (TR) and corona, and known as Dark Halos (DHs).  DHs have been insufficiently studied, and the mechanisms behind their darker emission remain unclear.}
   {This study aims to investigate for the first time the fine structure of a DH observed by the EUV High Resolution Imager (HRI$_{EUV}$) onboard the ESA's Solar Orbiter (SO) mission and its appearance in the TR.}
   {We utilized the extensive 1-hour dataset from SO on 19 March 2022, which includes high-resolution observations of NOAA 12967 and part of the surrounding DH. We analyzed the dynamics of the HRI$_{EUV}$ DH fine structure and its appearance in the HRI$_{Ly\alpha}$ image and the Spectral Imaging of the Coronal Environment (SPICE) Ly$\beta$, \ion{C}{III}, \ion{N}{VI}, \ion{O}{VI} and \ion{Ne}{VIII} lines, which sample the TR in the logT (K) $\sim$ 4.0 - 5.8 range. This analysis was complemented with a simultaneous B$_{LOS}$ magnetogram taken by the High Resolution Telescope (HRT).}
   {We report the presence of a peculiar fine structure which is not observed in the quiet Sun, characterized by combined bright EUV bundles and dark regions, arranged and interconnected in such a way that they cannot be clearly separated. They form a spatial continuum extending approximately radially from the AR core, suggesting a deep connection between the DH and the AR. Additionally, we find that the bright EUV bundles are observed in all the SPICE TR lines and the HRI$_{Ly\alpha}$ band and present photospheric B$_{LOS}$ footprints in the HRT magnetogram. This spatial correlation indicates that the origin of the 174 \r{A} DH may lie in the low atmosphere, i.e. photosphere/chromosphere.}
   {}

   \titlerunning{First Solar Orbiter observation of a dark halo in the solar atmosphere}
   \authorrunning{Lezzi, S.M., et al.}
   \maketitle
%

\section{Introduction}

\begin{figure*}
    \centering
    \includegraphics[scale=.66,trim= 10 245 25 200,clip]{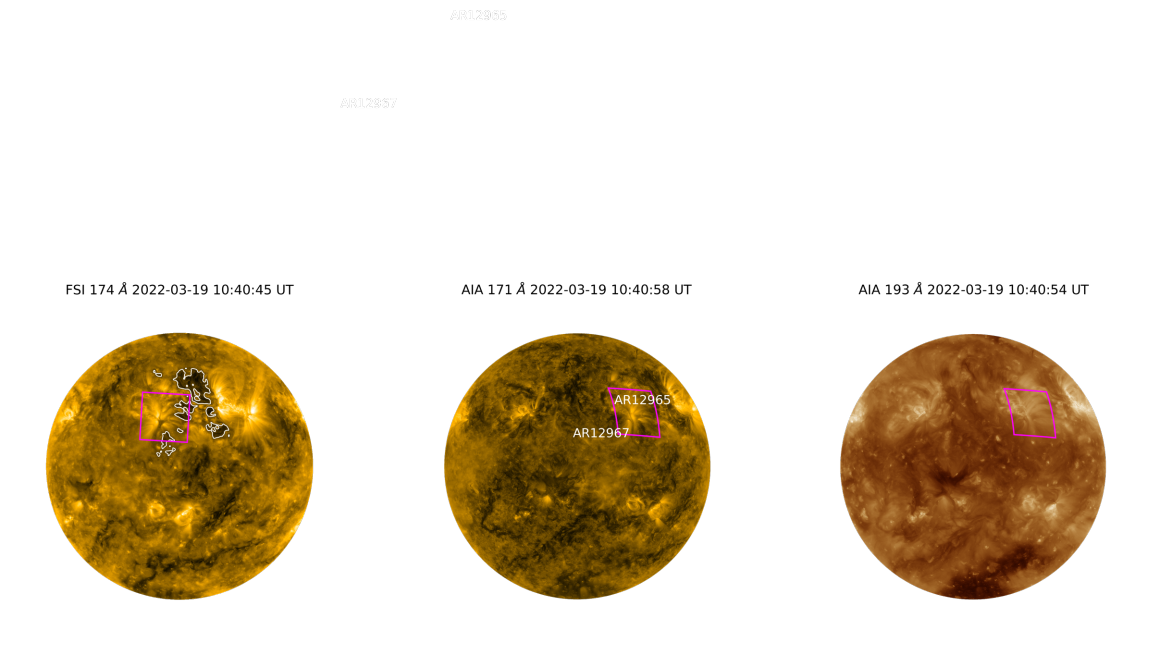}
    \includegraphics[scale=.64,trim= 10 30 25 235,clip]{fig1_2.png}
    \caption{EUI/FSI 174 \r{A}, AIA 171 \r{A} and 193 \r{A} full-disk images from March 19, 2022. Since SO was not on the Sun-Earth line during these observations, the FSI and AIA images show different portions of the disk. HRI$_{EUV}$ FoV is represented by the magenta box. It partially covers the DH (white contour) which fills the space between NOAA 12967 and NOAA 12965 and is presumably created by their presence. The white contour is obtained through an intensity threshold set to the 57\% of the mean FSI 174 \r{A} intensity of the disk, following what has been done in Paper I on the AIA 171 \r{A} full-disk image.   
    }
    \label{fig1}
\end{figure*}

Dark halos (DHs; \citealt{lezzi2023}, hereafter Paper I, and references therein) are regions characterized by reduced emission compared to the quiet Sun observed around solar active regions (ARs) at various wavelengths and wavebands covering the chromosphere, transition region (TR) and corona. DHs are usually large-scale structures which are difficult to observe in their entirety, except with instruments such as the Atmospheric Imaging Assembly (AIA; \citealt{lemen2012}) on board the Solar Dynamics Observatory (SDO; \citealt{pesnell2012}). However, SDO has limited TR coverage and does not provide spectral information. For this reason they have been poorly studied in the past and their origin, global structure and evolution are still unknown. 

The DHs observed in the AIA 171 \r{A} band have been shown to influence the total solar EUV irradiance. Indeed, \citet{toriumi2020} found that during extremely quiet solar conditions the 171 \r{A} disk-integrated irradiance decreases from the quiet Sun level when transiting events (such as an isolated sunspot, a spot-less plage, and emerging flux) cross the disk and the DH is contributing to the disk emission more than the AR. Moreover, as for instance \citet{wang2011}, \citet{singh2021} and Paper I show, DHs are normally found in the proximity of ARs. This coexistence suggests that they are intimately related to the global AR’s magnetic configuration and therefore unveiling the specific physical mechanism causing their darker appearance is of crucial importance in order to completely understand the heating mechanisms of ARs and of the solar atmosphere itself.  

Recently, in Paper I the authors took advantage of the full-disk mosaic observing mode of the Interface Region Imaging Spectrograph (IRIS; \citealt{depontieu2014}) to study the emission properties of the DH around NOAA 12706 by combining IRIS chromospheric and TR lines with SDO/AIA filtergrams and SDO/Helioseismic and Magnetic Imager (HMI;\citealt{schou2012}) magnetograms.  
In line with studies conducted by previous researchers (e.g. \citealt{rutten2007}; \citealt{pietarila2009}), the authors observed a fibril vortex in the chromosphere surrounding NOAA 12706. This structure, termed in Paper I “chromospheric fibrillar DH”, was most distinctly observed in the \ion{Mg}{II} h$_3$ and k$_3$ mosaics. They also noticed a corresponding dark shadow in the chromospheric/TR \ion{C}{II} and \ion{Si}{IV} mosaics, coinciding with the region covered by \ion{Mg}{II} fibrils, in which instead no clear fibrillar pattern could be detected. Instead, in the AIA bands they found a DH of larger extent and most clearly observed in the AIA 171 \r{A} band (as also reported by \citealt{wang2011}; \citealt{singh2021}) which they referred to as “coronal DH”. The relationship between the chromospheric (fibrillar) and the coronal (mainly 171 \r{A}) DHs still needs to be clarified and high-resolution co-spatial chromospheric, TR and coronal observations of the region shared by both the chromospheric fibrillar and the 171 \r{A} DHs are expected to provide useful insights.  


In this work we present the first analysis of observations of a DH in the solar atmosphere by instruments operating on the ESA's Solar Orbiter (SO; \citealt{muller2020}) mission. The data were taken on March 19, 2022 by EUI (Extreme Ultraviolet Imager;  \citealt{rochus2020}), SPICE (Spectral Imaging of the Coronal Environment; \citealt{spiceconsortium}) and PHI (Polarimetric and Helioseismic Imager; \citealt{solanki2020}). 

We take advantage of the SO proximity to the Sun and the high-resolution capabilities of the HRI$_{EUV}$ telescope to study the DH observed in the 174 \r{A} waveband (similar to the AIA 171 \r{A}) and investigate its fine structure with unprecedented spatial and temporal resolution. The details of this fine structure are the focus of this study. Furthermore, SPICE’s large temperature coverage provides insights in a wide range of TR lines (logT[K] $\sim$ 4.0 - 5.8), sampling temperatures from the chromosphere almost to the corona that were not probed in Paper I. In addition, the PHI High Resolution Telescope (HRT; \citealt{gandorfer2018}) complements the dataset by providing insight into the photospheric magnetic field of the area. Finally, we discuss a plausible physical scenario related to the influence of the AR’s magnetic field which may explain their observational properties of the DH.

\section{Observations}

The data we use in this study come from the SO remote sensing instruments EUI, SPICE and PHI that on March 19, 2022 were pointing at NOAA 12967, a decaying AR with an $\alpha$ configuration, as part of the SOOP (SO Observing Plan) campaign L\_SMALL\_HRES\_CAD\_Slow-Wind-Connection (described in section 4.14 of \citealt{zouganelis2020}). On that day SO, which was approaching its fourth perihelion, was at a distance of 0.36 AU from the Sun and had an angular separation of 36° west in Carrington longitude with the Earth-Sun line. From the Earth/SDO point of view, NOAA 12967 rotated onto the solar disk on March 12 and disappeared on March 22, and was accompanied on its West side by the more complex ($\beta\gamma$ configuration) AR NOAA 12965, which has been extensively studied (e.g., \citealt{berghmans2023}; \citealt{chitta2022}; \citealt{mandal2023a}; \citealt{mandal2023b}).  

Fig. \ref{fig1} shows the full-disk images in the EUI Full Sun Imager (FSI) 174 \r{A}, AIA 171 \r{A} and AIA 193 \r{A} channels taken on March 19, 2022, at 10:40:45 UT, 10:40:58 UT and 10:40:54 UT, respectively.
EUI's FoV, shown as a magenta box on the FSI image, falls partially on both NOAA 12967 and, in the top right region of the FoV, in the dark area observed between NOAA 12967 and NOAA 12965, which in the AIA 171 \r{A} image is instead covered by the AR loop emission. We consider this area of reduced emission compared to the quiet Sun, identified by the white contour of Fig. \ref{fig1} (left panel), as the DH presumably created and shared by NOAA 12967 and NOAA 12965. Paper I showed that, contrarily to coronal holes, DHs are clearly outlined in the AIA 171 \r{A} waveband, while they are nearly indistinguishable from quiescent coronal areas in the AIA 193 \r{A} waveband. As shown in Fig. \ref{fig1}, our region of interest is dark in the AIA 171 \r{A} but not in the 193 \r{A} band. Therefore, we are confident that it is the DH originated in the atmosphere by the presence of both ARs.

Fig. \ref{hrieuvfov} shows the HRI$_{EUV}$ snapshot taken at 10:36:00 UT with the co-aligned FoVs of SPICE (red box) and HRT (blue box) superimposed. More information about the alignment process of the images of the different instruments is described in the Appendix.

\subsection{HRI data}
HRI$_{EUV}$ dataset (data release 6.0\footnote{\url{https://www.astro.oma.be/doi/ROB-SIDC-SolO_EUI-DataRelease6.0_2023-01.html}}) consists of high-resolution images at 5 s fast cadence taken between 10:36:00 UT and 11:35:55 UT. The HRI$_{EUV}$ telescope has the peak of its thermal response function at about 1 MK, due to the emission of \ion{Fe}{IX} and \ion{Fe}{X} (spectral lines at 171.1 \r{A} and at 174.5 \r{A} and 177.2 \r{A}, respectively). 

Since HRI$_{EUV}$ has a PSF with a FWHM approximately twice the image pixel scale of 0.492$\arcsec$ pixel$^{-1}$, the spatial resolution of these images is $\sim$258 km on the Sun, which is a factor $\sim$ 2.8 higher than AIA's spatial resolution. The temporal resolution of HRI$_{EUV}$ is also higher than that of AIA, by a factor of 2.4. 

The HRI$_{EUV}$ observing sequence includes jitter, which we removed by following the method described by \citet{chitta2022}. We complement the HRI$_{EUV}$ view of the region by exploiting the image taken on the same day at 10:36:00 UT by the HRI$_{Ly\alpha}$ telescope, that observes the Lyman-$\alpha$ waveband centred at 121.6 nm, corresponding to the hydrogen Lyman-$\alpha$ (\ion{H}{I} Ly${\alpha}$), the strongest line in the solar ultraviolet spectrum (\citealt{curdt2001}). This line is optically thick (\citealt{vial1982}; \citealt{woods1995}) and formed mostly in the chromosphere and TR (e.g., \citealt{vernazza1981}; \citealt{vourlidas2010}).  

\subsection{SPICE data}
\begin{figure}
    \includegraphics[scale=.69,trim= 14 10 11 10,clip]{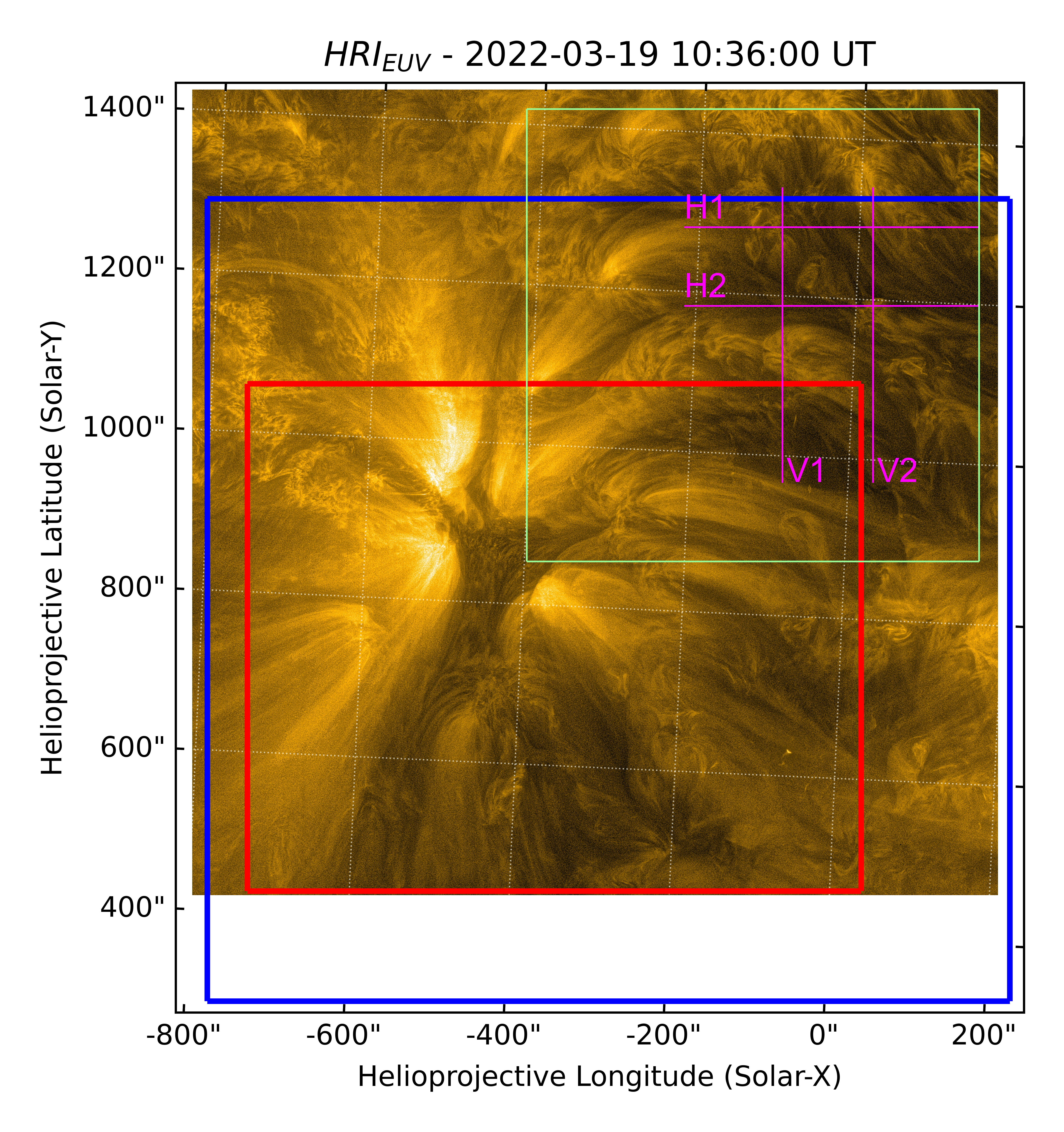}
    \caption{EUI/HRI$_{EUV}$ snapshot taken on March 19, 2022 at 10:36:09 UT. The red box is the FoV of the SPICE co-aligned image taken at 10:37:00 UT; the blue box is the FoV of the HRT co-aligned image taken at 10:36:09 UT. The green box represents the close-up view of DH fine structure shown in Fig. \ref{qsfig}. The magenta horizontal (H1 and H2) and vertical (V1 and V2) lines are the slices chosen to build the slice plots shown in Fig. \ref{iplots}. The image is enhanced with the Multi-scale Gaussian normalization algorithm (\citealt{morgan2014}).}
    \label{hrieuvfov}
\end{figure}

The SPICE data we use (data release 4.0\footnote{\url{https://spice.osups.universite-paris-saclay.fr/spice-data/release-4.0/release-notes.html}}) is the 192-step raster taken on March 19, 2022 at 10:37:00 UT. It has a duration of 96 min 48 s and an exposure time of 30 s. The selected slit is 4$\arcsec$ wide and the field of view (FoV) is smaller than HRI$_{EUV}$'s (see Fig. \ref{hrieuvfov}). 

Among the available spectral lines, we select those that are not blended: \ion{H}{I} Ly$\beta$ at 1025.72 \r{A} (log $T_e$[K]  = 4.0), \ion{C}{III} at 977.03 \r{A} (log $T_e$[K]  = 4.8), \ion{N}{IV} at 765.2 \r{A} (log $T_e$[K]  = 5.08), \ion{O}{VI} at 1031.93 \r{A} (log $T_e$[K]  = 5.5) and \ion{Ne}{VIII} at 770.42 \r{A} (log $T_e$[K]  = 5.8). The FoV of the \ion{N}{VI} and \ion{Ne}{VIII} rasters is 198 $\times$ 157 Mm, while it is 198 $\times$ 164 Mm for the \ion{C}{III}, \ion{Ly}{$\beta$} and \ion{O}{VI} spectral lines. This is due to the slit being a bit shorter in the SW detector than in the LW, with 628 and 600 pixels respectively before the trimming of the FoV (see Fig. 4 of \citealt{spiceconsortium}).

\subsection{HRT data}
To investigate the photospheric magnetic field underlying the DH fine structure, we exploit the cotemporal and cospatial B$_{LOS}$ magnetogram taken by the PHI/HRT telescope on March 19, 2022 at 10:36:09 UT (data release 1\footnote{\url{https://www.mps.mpg.de/solar-physics/solar-orbiter-phi/data-releases}.}).


\begin{figure*}
    \centering
    \includegraphics[scale=.6,trim= 0 10 0 10,clip]{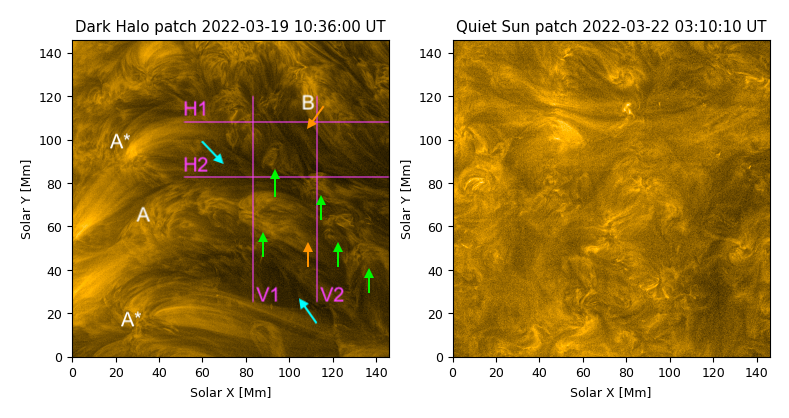}
    \includegraphics[scale=.407,trim= 0 0 0 0,clip]{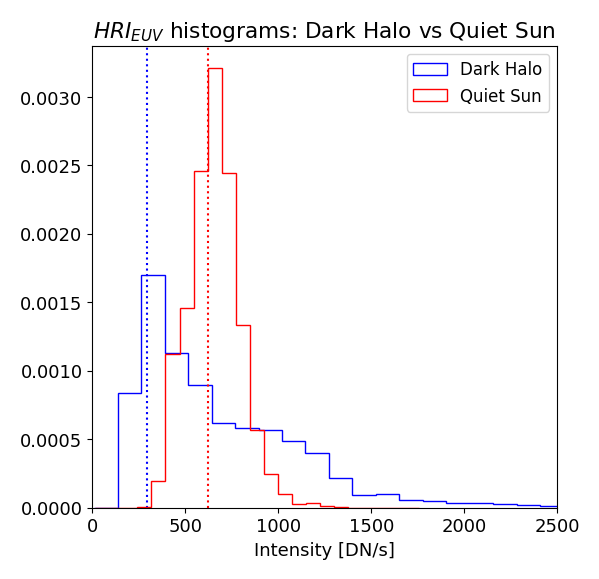}
    \caption{\textit{Left and middle panels}: HRI$_{EUV}$ close-up view of the DH fine structure taken from Fig. \ref{hrieuvfov} and quiet Sun patch of the same size taken on March 22, 2022. The size of both patches is 146 $\times$ 146 Mm. The images are enhanced with the Multi-scale Gaussian normalization algorithm (\citealt{morgan2014}). The ‘A’ mark labels a regular EUV bundle observed in the DH fine structure; the ‘A*’ mark labels a comet-like EUV bundle observed in the DH fine structure. The green arrows point at some examples of dome-like bundles; the orange arrow points at the darker filamentary body associated with the dome located at (80,65) Mm. The cyan arrows point to the dark areas that cannot be easily traced back to any bundle. The magenta horizontal (H1 and H2) and vertical (V1 and V2) lines are the slices chosen to build the slice plots shown in Fig. \ref{iplots}. See Section 3 for more information.  \textit{Right panel}: Normalized histograms of the DH fine structure and quiet Sun patches shown in the left and middle panels. The red and blue dotted lines represent the mode of the QS and DH distributions, respectively.}
    \label{qsfig}
\end{figure*}

\begin{figure}
    \centering
    \includegraphics[scale=.38,trim= 0 58 40 40,clip]{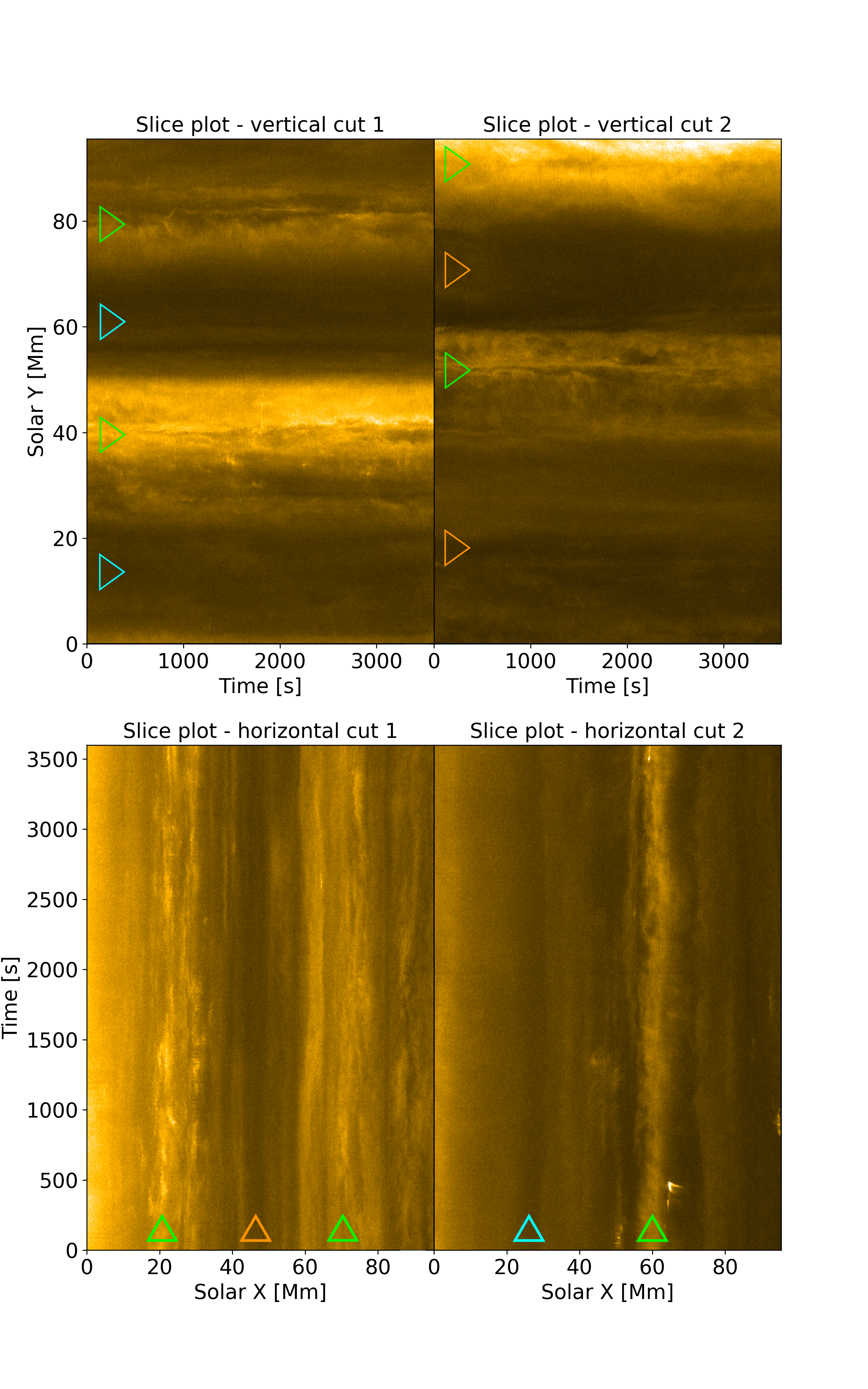}
    \caption{HRI$_{EUV}$ slice plots of the vertical (top) and horizontal (bottom) slices shown in Figs. \ref{hrieuvfov} and \ref{qsfig}. The green triangles are located on the bulk of bright EUV bundles; the orange triangles are located on dark regions that can be traced back to EUV bundles in the HRI$_{EUV}$ image; the cyan triangle is located on the bulk of a dark region that cannot be traced back to any EUV bundle in the HRI$_{EUV}$ image.} 
    \label{iplots}
\end{figure}

\begin{figure}
    \centering
    \includegraphics[scale=.52,trim= 14 15 30 13,clip]{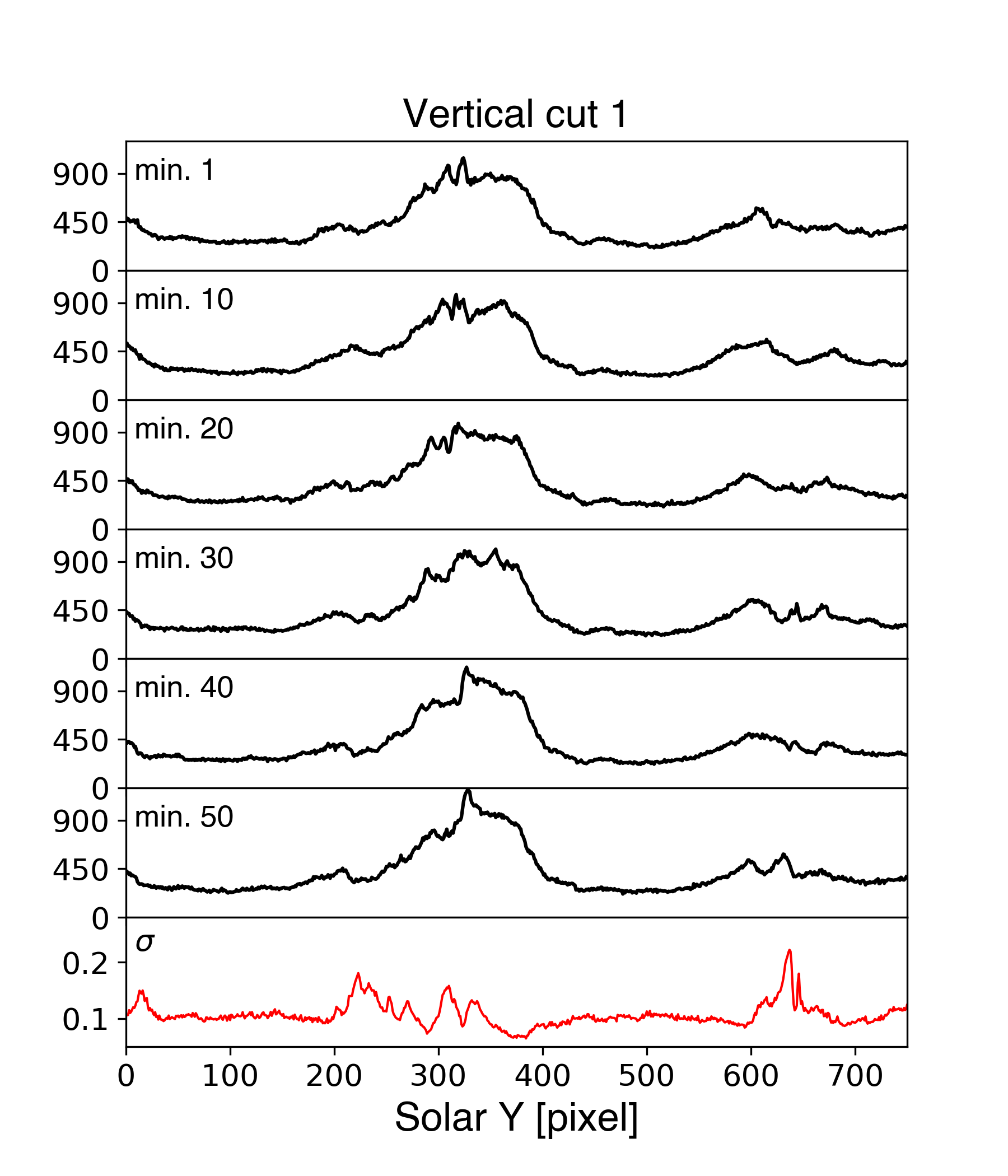}
    \caption{Intensity fluctuations (in DN/s) of the vertical cut V1 shown in Figs. \ref{hrieuvfov} and \ref{qsfig}. The intensity profiles (black lines) are 1-min averages and are shown at a 10-min interval. The red line is the standard deviation (normalized to the mean) of the cut for the 1-hour HRI$_{EUV}$ timeseries.} 
    \label{iprofiles}
\end{figure}

\begin{figure*}
    \centering
    \includegraphics[scale=.75,trim= 0 108 0 90,clip]{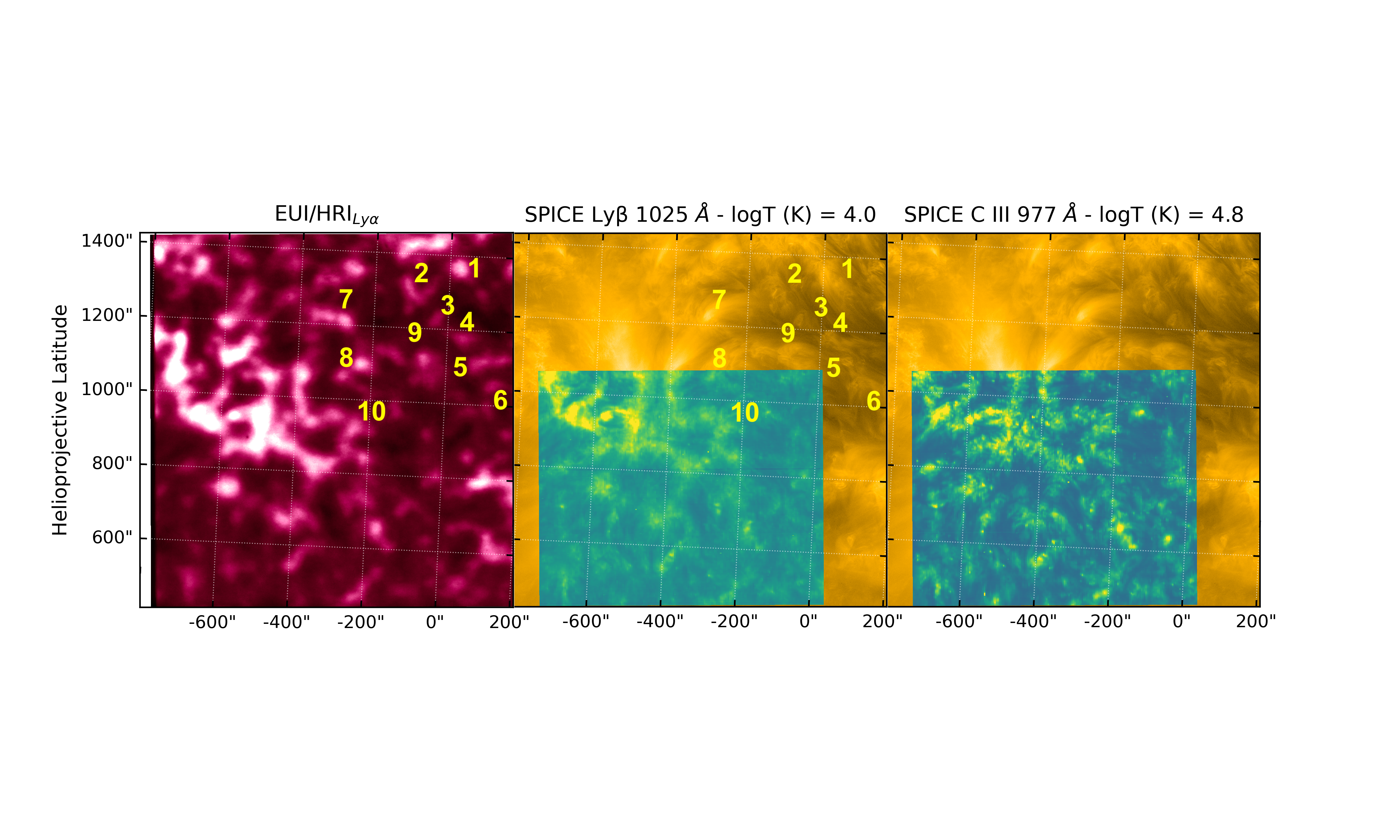}
    \includegraphics[scale=.75,trim= 0 90 0 90,clip]{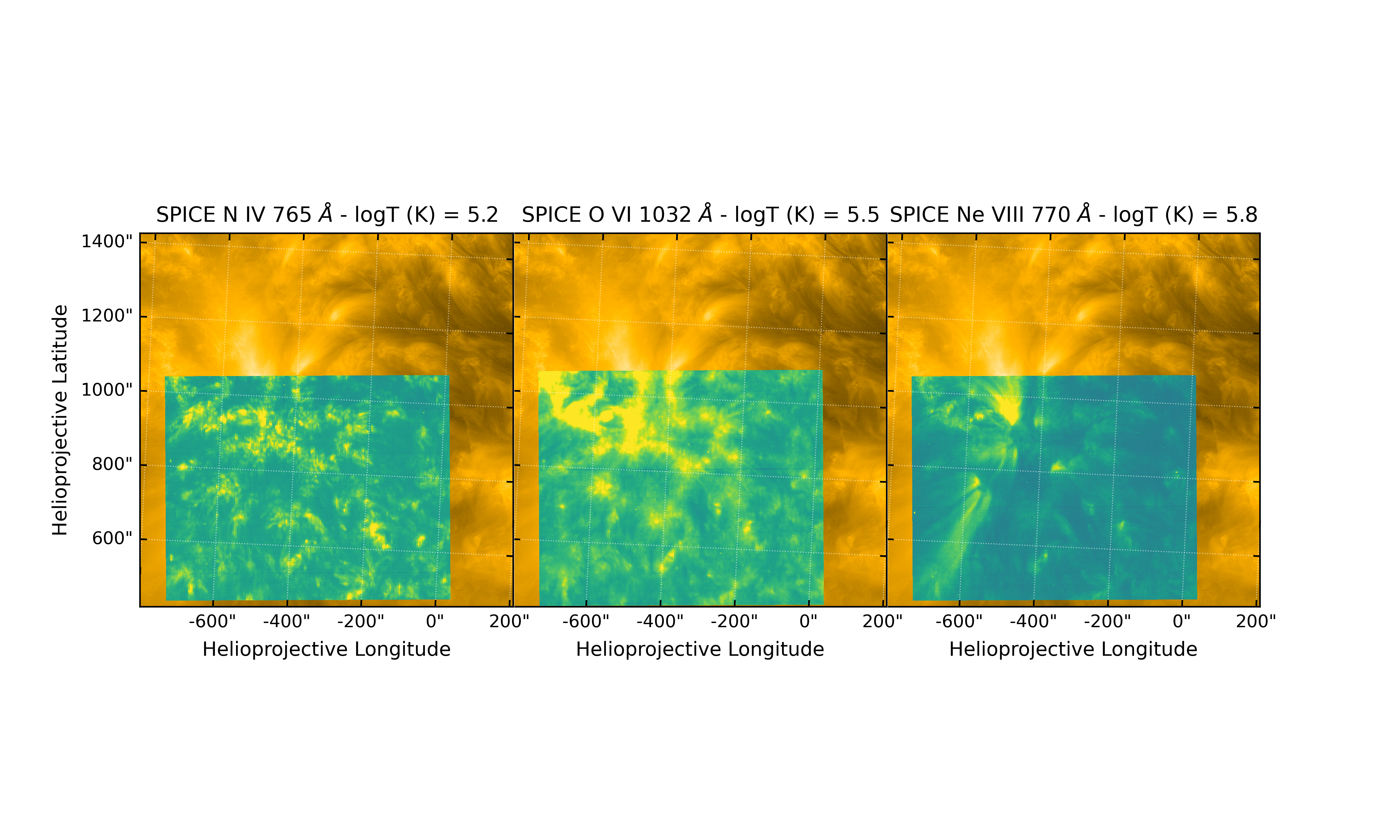}
    \caption{Co-aligned HRI$_{Ly\alpha}$ image and SPICE integrated intensities showing the chromospheric/TR view of NOAA 12967 and the DH fine structure. The SPICE observations in the H-Ly$\beta$, \ion{C}{III}, \ion{N}{IV}, \ion{O}{VI} and \ion{Ne}{VIII} lines (ordered as function of temperature) are overplotted on the HRI$_{EUV}$ image. The numbers in the top left and middle panels mark the locations of the EUV bundles shown in Fig. \ref{qsfig} that have a clear counterpart in the HRI$_{Ly\alpha}$ image. The bundle marked by number 10 also has counterparts observed in the SPICE rasters.}
    \label{lya_and_spice}
\end{figure*}

\begin{figure*}
    \centering
    \includegraphics[scale=.75,trim= 10 10 10 10,clip]{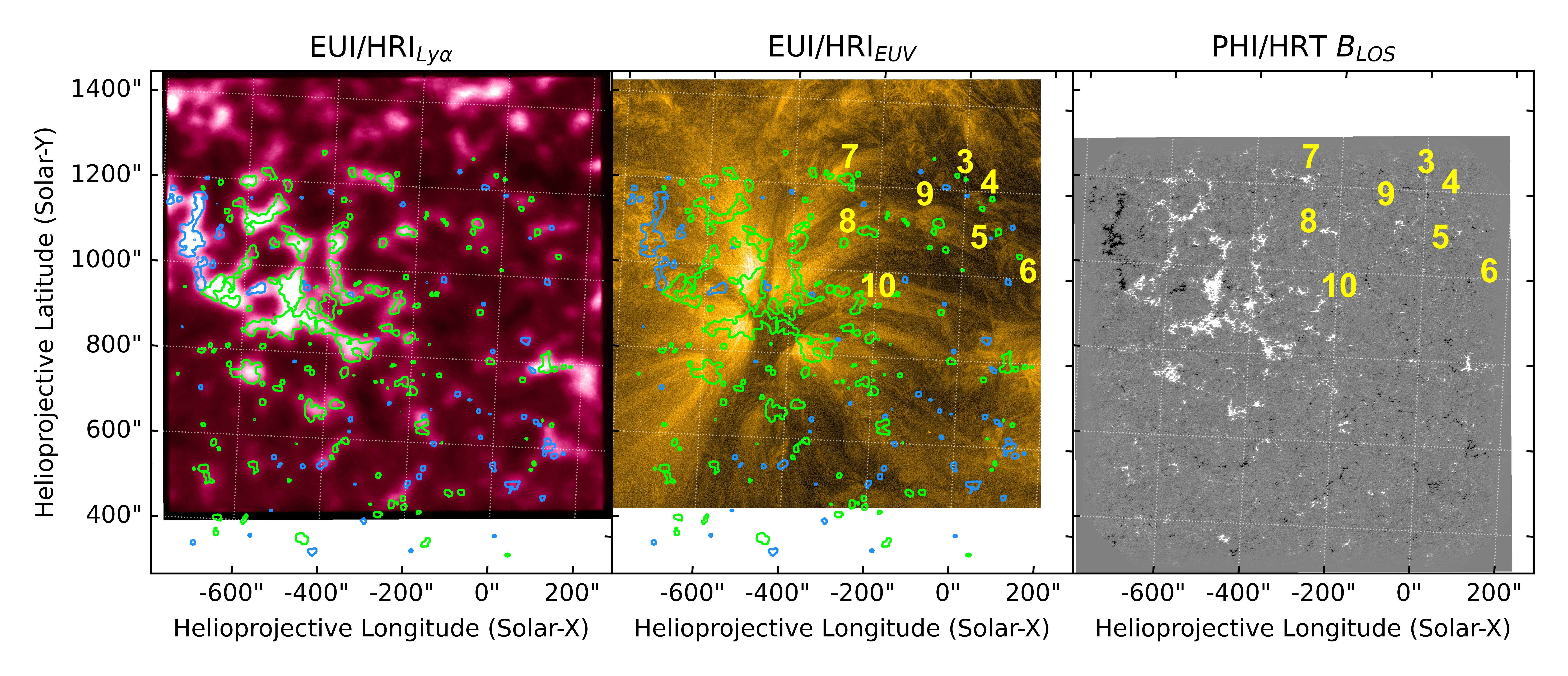}
    \caption{Overview of EUI/HRI and PHI/HRT observations from March 19, 2022: HRI$_{Ly\alpha}$ image (\textit{left}), HRI$_{EUV}$ image (\textit{middle}) and HRT B$_{LOS}$ magnetogram (\textit{right}) taken at 10:36:00 UT, 10:36:09 UT and 10:36:09 UT, respectively. The HRI$_{EUV}$ image is enhanced with the Multi-scale Gaussian normalization algorithm (\citealt{morgan2014}). The numbers (same of Fig. \ref{lya_and_spice}) label the EUV bundles that have bright counterparts in the HRI$_{Ly\alpha}$ image. Labels 1 and 2 have not been displayed since they are located outside the HRT FoV. The green (positive polarity) and cyan (negative polarity) contours enclose the regions where |B$_{LOS}$| > 16.6 G. Find more details in Section \ref{loweratm}.}
    \label{phi}
\end{figure*}

\section{The EUV fine structure of the Dark Halo}

HRI$_{EUV}$ images show that the 174 \r{A} DH has a detailed fine structure that is remarkably different from the one typically seen in the quiet Sun. To prove this, in Fig. \ref{qsfig} we show a 146 $\times$ 146 Mm HRI$_{EUV}$ close-up view of DH fine structure (corresponding to the green box of Fig. \ref{hrieuvfov}) and a quiet Sun patch of the same size taken at disk centre on March 22, 2022 at 03:10:10 UT. The DH patch of Fig. \ref{qsfig} (left panel) exhibits a fine structure characterized by the juxtaposition of dark regions and bright EUV bundles, each assuming diverse and complex appearances:
\begin{itemize}
    \item The bright EUV bundles, which display an orientation approximately radial with respect to NOAA 12967, share similarities, exhibiting nonetheless subtle differences in appearance that prompt an attempt at a tentative finer classification. Some bundles seem to originate from the AR itself (see e.g. the ‘A’-marked group of EUV bundles in Fig. \ref{qsfig}). Some of them exhibit a “comet-like” fan shape on top (cases ‘A*’ in Fig. \ref{qsfig}). Part of them present a “dome-like” bright part (e.g. see green arrows in Fig. \ref{qsfig}) which in certain instances extends into a darker filamentary body (orange arrows in Fig. \ref{qsfig}). In one specific case the bundle appears to contain intrinsically dark streaks and veins (‘B’ mark in Fig. \ref{qsfig}). 
     
    \item The other defining feature of the DH fine structure comprises dark regions which often appear interconnected and tangled to the EUV bundles, in such a way that it is difficult to clearly separate them. Moreover, these dark regions are more or less distinctly associated with nearby EUV bundles. For instance, dark filamentary channels (see orange arrows in Fig. \ref{qsfig}) can be associated to the bright bundles located at $\sim$ (85,70) Mm and (110,130) Mm in Fig. \ref{qsfig}. However, other dark regions (highlighted by cyan arrows in Fig. \ref{qsfig}) appear not to be morphologically connected to any specific bundle.  
\end{itemize} 

To study the “coronal” (AIA 171 \r{A}) DH, in Paper I the authors used a delimiting contour obtained via an intensity threshold (equal to 57\% of the mean value of the disk intensity) specifically chosen to match the dark area seen by eye in the AIA 171 \r{A} image. This HRI$_{EUV}$ close-up view shows instead that it is not possible to treat the 174 \r{A} DH and the AR separately due to their coupling in terms of fine structure, i.e. dark filamentary channels proceeding from bright AR-related EUV bundles. EUV bundles and dark channels form a spatial continuum in which they are arranged in an interconnected and blended pattern. As a consequence, we point out that the method of the intensity threshold first adopted in Paper I to identify and study the 171/174 \r{A} DHs has not to be meant as a strict definition of the spatial extent of such structures, which are essentially entangled with and inseparable from the corresponding ARs. Contour thresholds are a convenient tool to localize DHs in the vicinity of ARs, especially in full-disk images, as discussed in Paper I. However, the results presented in this work indicate that, in reality, the transition between a DH and the associated AR does not present any sharp boundary. Contours remain useful, instead, to outline the outer edge of a DH.

On the other hand, the quiet Sun (see Fig. \ref{qsfig}, middle panel) exhibits the typical fuzzy configuration due to the magnetic field starting to diverge at these higher temperatures ($\sim$1 MK) and heights (\citealt{fludra2021}). In Fig. \ref{qsfig} (right panel) we also show the normalized histograms of the entire DH and quiet Sun patches which are shown in the left and middle panels of Fig. \ref{qsfig}. The quiet Sun has a single peak gaussian distribution, with a mode value of 625 DN/s (red dotted line); the DH shows a single peak non-gaussian distribution with a wide right wing, representative of the contributions from the bright EUV bundles and the dark regions, associated with the mode of 297 DN/s (blue dotted line). This suggests that the overall emission observed in the 174 \r{A} DH is reduced compared to the quiet Sun because of its particular configuration characterized by a blend of bright EUV bundles and dark regions.

\subsection{Intensity fluctuations}
We exploit the HRI$_{EUV}$ 5 s cadence dataset to study the intensity fluctuations of the 174 \r{A} DH fine structure. We analyze the time evolution of the HRI$_{EUV}$ intensity of four virtual slices, two vertical and two horizontal (marked respectively as V1, V2, and H1, H2 in Fig. \ref{qsfig}), taken inside the DH and shown in Figs. \ref{hrieuvfov} and \ref{qsfig} (left panel) and one HRI$_{EUV}$-pixel wide.
Different slices have been tested and the four shown here have been chosen to best sample both bright bundles and dark regions. In particular: V1 falls on two EUV bundles and two dark regions apparently not associated to any bundle; V2 falls on two EUV bundles and two dark regions associated with them; H1 samples two EUV bundles and one dark region associated to the bundle on the right; H2 samples a dark region not associated to any bundle and a bright bundle. We point out that there is no sharp separation between the observed bundles and dark regions since they are combined in the fine structure of the DH.  
We track in time each slice by building a slice plot, which displays the intensities along the slice as functions of time (Fig. \ref{iplots}).
In addition, we present the slices' intensity profiles (depicted as black lines in Figs. \ref{iprofiles} and \ref{iprofiles2}) at intervals of ten minutes, averaged over the initial minute, to illustrate the fluctuation in intensity across various regions of interest. To measure the extent of intensity variability, we display, for each pixel along the slice, the standard deviation normalized to the mean (illustrated as a red line in Figs. \ref{iprofiles} and \ref{iprofiles2}), calculated based on the entire dataset spanning one hour.  

Both the slice plots (Fig. \ref{iplots}) and the intensity profiles (Figs. \ref{iprofiles} and \ref{iprofiles2}) show that the DH fine structure is quite stable, with temporal fluctuations of $\sim$ 10\%, as shown by the standard deviations normalized to the mean (Figs. \ref{iprofiles} and \ref{iprofiles2}). Moreover, the bright bundles show slightly higher variability, with fluctuations reaching peaks of 20\% in one hour. We point out that from the intensity variability no distinction seems to rise between the dark regions associated (Fig. \ref{iplots}, orange triangles) and not associated (Fig. \ref{iplots}, cyan triangle) with EUV bundles.  

\subsection{The link to the lower atmosphere}\label{loweratm}

In Paper I the authors used the IRIS \ion{Mg}{II}, \ion{C}{II} and \ion{Si}{IV} lines to show the \ion{Mg}{II} chromospheric fibrillar DH and its counterparts at $\sim$25,000 K and $\sim$60,000 K, respectively. For this particular target no relevant IRIS data were available and therefore it was not possible to exploit those lines. However, the SPICE images show the TR counterparts of a 174 \r{A} DH in a larger TR temperature range (logT[K] $\sim$ 4.0 - 5.8).  Therefore we complement the HRI$_{EUV}$ dataset with co-aligned chromospheric/TR images of SPICE and HRI$_{Ly\alpha}$ and the photospheric HRT B$_{LOS}$ magnetogram, which together provide insights on the origin of the DH emission. In Fig. \ref{lya_and_spice} the SPICE rasters (in which intensities are obtained by summing over each spectral pixel window) are ordered as a function of temperature (from the coolest to the hottest) and show the emission response to the temperature increase in the atmosphere, with the cooler lines, H-Ly$\beta$ and \ion{C}{III}, displaying the chromospheric/TR view of the AR plage, and the hottest line, \ion{Ne}{VIII}, resembling the appearance of HRI$_{EUV}$ 174 \r{A} image. Moreover, since the H-Ly$\alpha$ line is formed mostly in the chromosphere and TR (e.g., \citealt{vernazza1981}), the HRI$_{Ly\alpha}$ image provides an extension of the limited SPICE FoV. 
 
We find that the EUV bundles observed by HRI$_{EUV}$, their brighter parts in particular, have a bright matching counterpart in the HRI$_{Ly\alpha}$ and SPICE (when the FoV allows it) images. In Fig. \ref{lya_and_spice} we show ten examples of HRI$_{EUV}$ bundles that we were able to trace back to co-spatial chromospheric/TR HRI$_{Ly\alpha}$ bright counterparts (see number labels in the top left and top middle panels of Fig. \ref{lya_and_spice}). The SPICE rasters, despite the smaller FoV, show this spatial correspondence as they exhibit the same bright patches observed in the Ly$\alpha$ and located at approximately the same positions of the 174 \r{A} EUV bundles (see e.g. label 10 in Fig. \ref{lya_and_spice}).

Fig. \ref{phi} shows the HRT B$_{LOS}$ magnetogram together with the HRI$_{EUV}$ and HRI$_{Ly\alpha}$ images. The superimposed contour encloses the photospheric magnetic concentrations (green for the positive polarity, cyan for the negative polarity) with |B$_{LOS}$| > 16.6 G. This value corresponds to twice the maximum value found by \citet{sinjan2022} for the noise level of the HRT B$_{LOS}$ magnetograms. This figure shows that the HRI$_{EUV}$ bundles seem to also match magnetic patches (of mainly positive polarity) in the HRT B$_{LOS}$ magnetogram. In particular, the bigger EUV bundles (labels 7, 8, 9 and 10) are evidently located over magnetic concentrations of positive polarity, while few negative polarity magnetic patches of reduced dimension can also be identified in the proximities of these bundles.
The association to the photospheric magnetic activity is instead more difficult and uncertain for the smaller EUV bundles (labels 3, 4, 5 and 6) because the observed magnetic patches (of both polarities) are of overall smaller dimensions and found not over but nearby the bundles. The visibility of the smaller magnetic concentrations in the HRT B$_{LOS}$ magnetogram is influenced by the sensitivity of the instrument, which has noise levels between 6.6 and 8.3 G (\citealt{sinjan2022}). The EUV bundles could occur between two main magnetic footpoints or be confined between bipolar magnetic features that in this case are not well resolved.  
On the other hand, the imperfect overlap of some of the magnetic concentrations on the EUV bundles could be a consequence of the expansion of the magnetic field with height in the atmosphere. \citet{jafarzadeh2017} present a Bifrost (\citealt{gudiksen2011}) simulation of two opposite-polarity (network) patches in the entire atmosphere which clearly visualize this idea (in a volume horizontally large 24 $\times$ 24 Mm$^2$, similar to the extent of the EUV bundles; see Fig. \ref{qsfig}). At a photospheric height $z_{ph}$, a magnetic field line touches the horizontal plane $P_{ph}$ at a point of coordinates (x$_{ph}$, y$_{ph}$); going up in the TR, the same magnetic field line touches the horizontal plane $P_{TR}$  located at $z_{TR}$ at a new point of coordinates (x$_{TR}$, y$_{TR}$). Consequently, if the EUV bundles are phenomena related to the presence of photospheric magnetic activity, as the spatial correlation in the B$_{LOS}$ magnetogram seems to suggest, then in the TR they are expected to occur at locations slightly different than those of the photospheric magnetic patches. 
Therefore, the overlap observed between the EUV bundles and the nearby magnetic concentrations is likely the combination of both limited instrument sensitivity and expansion of the magnetic field lines in the atmosphere. However, some of the small magnetic concentrations found in the neighborhoods of the EUV bundles may as well not be related to them. A similar scenario has been found by \citet{kahil2020} for a campfire (\citealt{berghmans2021}) dataset: these authors showed that about a quarter of the campfires was observed over small scattered magnetic features which could not be identified as magnetic footpoints of the campfire hosting loops. Even if this also applied to the DH fine structure, the bright EUV bundles are observed in all the SPICE TR lines and the HRI$_{Ly\alpha}$ band and present (at least for the bigger EUV bundles) photospheric B$_{LOS}$ footprints in the HRT magnetogram. These spatial correlations,  combined with the coupling between EUV bundles and dark channels, suggests that the origin of the 174 \r{A} DH lies in the lower atmosphere, specifically in the photosphere/chromosphere.
Consequently, these findings suggest that the “coronal DH” term proposed in Paper I more precisely refer to a TR/lower coronal phenomenon. For simplicity, we would still retain the “coronal DH” term or adopt as an alternative, “174 \r{A} DH”.

\section{Discussion and conclusions}
This SO dataset represents a unique opportunity to look into the fine structure of the 171/174 \r{A} DH, since it comprises high-cadence high-resolution HRI$_{EUV}$ images complemented by the wide temperature coverage in the chromospheric/TR layers provided by SPICE and HRI$_{Ly\alpha}$ and the PHI/HRT photospheric LOS magnetic field.
We primarily focus our attention on the right-hand side of the HRI$_{EUV}$ image, which captures part of the DH shared by the two AR NOAA 12967 and NOAA 12965. 
We report a peculiar fine structure which is not observed in the quiet Sun (see Fig. \ref{qsfig}) and is characterized by combined brighter EUV bundles, some of them clearly related to the AR, and darker regions arranged to form a composite pattern. The bright EUV bundles and the darker features are intertwined in such a way that it is often  not possible to clearly distinguish and separate them. They constitute a spatial continuum that extends approximately radially from the AR core, implying a profound connection between the DH and the presence of the AR itself. 

The dark elements of the DH fine structure manifest in two distinct forms: as dark filamentary channels, often associated with EUV bundles; or as unmarked dark regions not clearly linked to EUV bundles. However, as depicted in Figs. \ref{iplots}, \ref{iprofiles} and \ref{iprofiles2}, the intensity fluctuations of these two types of dark regions exhibit similarities, suggesting that they may not be fundamentally distinct. It is plausible that the apparent differences arise from a signal too faint to be discerned by the instrument, particularly in the case of the unmarked dark areas. 

The EUV bundles, which represent the brighter component of the DH fine structure, have a counterpart in the HRI$_{Ly\alpha}$ band and in SPICE Ly$\beta$, \ion{C}{III}, \ion{N}{IV}, \ion{O}{VI} and \ion{Ne}{VIII} TR lines, corresponding to the  temperature range logT[K] $\sim$ 4.0 - 5.8. They also match B$_{LOS}$ patches in the HRT magnetogram. These spatial correlations suggest that the mechanism behind these features likely originates in the photosphere/chromosphere. In this context, it is essential to examine the connection between the 174 \r{A} DH and the chromospheric fibrillar DH, as observed, for instance, in Paper I around NOAA 12706 in the \ion{Mg}{II} h$_3$\&k$_3$ IRIS mosaics. In this regard, an overall similarity can be seen in the HRI$_{EUV}$ images between the EUV DH’s fine structure and the static low-lying chromospheric H$\alpha$/\ion{Ca}{II}/\ion{Mg}{II} fibrils normally arranged around AR (e.g. \citealt{rutten2007}; \citealt{kianfar2020}). See, for example, the EUV bundles at about (-450",680") and (-220”, 1100”), which resemble the “head-segment” part (\citealt{kianfar2020}) of fibrils located above some magnetic concentrations around the decaying AR shown in Fig. 1 (panels d, e, f) by \citet{kianfar2020}. In both the HRI$_{Ly\alpha}$ image and SPICE lines, despite the low spatial resolution of the images, corresponding cospatial dark areas are also found around the south part of the plage, which (in particular for the SPICE \ion{C}{III} line) recall the \ion{Mg}{II} chromospheric fibrillar DH and/or the \ion{C}{II}/\ion{Si}{IV} shadow of the chromospheric fibrillar DH observed in Paper I. However, it is also worth noting that the slice plots do not exhibit the parabolic tracks observed by \citet{mandal2023a} in the space-time diagram of the EUV signatures of dynamic fibrils (DFs; \citealt{depontieu2007}) located both close to the leading sunspot of the AR and in moss regions. Therefore the relationship between the 174 \r{A} and the chomospheric fibrillar DHs still remains unclear and further analysis including high-resolution observations of chromospheric lines (H$\alpha$, \ion{Mg}{II} or \ion{Ca}{II}) will be crucial. 

In any case, if the chromospheric fibrillar DH is indeed linked to the AR’s magnetic field (given that fibrils serve as indicators of the magnetic field topology), it seems likely that the occurrence of the 174 \r{A} DH, characterized by a mixture of bright EUV bundles and fainter dark regions, is also intimately related to the presence of the AR’s magnetic field. 

\citet{baker2023} used a potential field source surface (PFSS) extrapolation to investigate the large-scale coronal magnetic configuration of NOAA AR 12967 and identified low-lying closed loops connecting NOAA 12967 to NOAA 12965 just above the DH area. In this context, the EUV bundles could be interpreted as the TR counterparts of the chromospheric roots of low-lying horizontal loops  extending from the outermost regions of the plage and linking the plage’s photospheric magnetic concentrations to smaller magnetic patches of opposite polarity located in the surrounding vicinity. We point out that nearly horizontal cool low-lying loops have been studied initially by \citet{antiochos1986} and later by \citet{sasso2015}, but in the context of explaining the TR emission of the QS.

The DH’s emission properties, i.e. the faint emission of these low-lying loops in the parts furthest from the plage roots, may be the consequence of the different propagation of magnetoacoustic waves caused by the inclination of the magnetic field in this area. In fact, the DH may be the effect of a reduced heating associated with the so-called “magnetic shadows” (\citealt{jess2023} and references therein), global acoustic wave power suppressions observed in the presence of strong magnetic fields (e.g., plage regions, where a group of concentrated magnetic features reside). \citet{pietarila2013} showed that chromospheric H$\alpha$ intensity oscillatory power in the 3-minute period range is visibly reduced in circumfacular regions (i.e. chromospheric fibrillar DH) relative to the surrounding quiet Sun. Therefore, we speculate that the missing acoustic wave energy flux in the horizontal magnetic fields at the plage edges may cause less heating in the upper chromospere and the consequent DH’s reduced emission. Moreover, this missing acoustic flux could perhaps feed the flux of Alfvén waves in corona (e.g., \citealt{cally2022a}), through the fast-to-Alfvén (\citealt{cally2011}) and slow-to-Alfvén (\citealt{cally2022b}) mode conversion (\citealt{cally2008}) that magnetohydrodynamic (MHD) waves exhibit in continuously non-uniform plasmas (e. g. the corona). This argument needs to be explored by further investigation.

To understand the nature of the 174 \r{A} DH fine structure, observations specifically designed to study its reduced emission and its relationship with the chromospheric fibrillar DH are necessary. These observations should be coordinated among the solar high-resolution instruments (both space and ground based) and include spectroscopic data to extensively sample both chromospheric and TR temperatures. They should also specifically target these dark areas in a mosaic mode to cover their entire (or almost) spatial extent. A new SOOP (R\_BOTH\_Hres\_Lcad\_Dark-Halos) is being prepared to fulfil these requirements. Although the SO dataset exploited in this work was not part of a campaign designed to study DHs, it has nevertheless produced interesting insights and therefore we expect this new SOOP to produce datasets that will definitively shed further light on the nature of DHs.

\noindent
  

\begin{acknowledgements}
        Solar Orbiter is a space mission of international collaboration between ESA and NASA, operated by ESA. The EUI instrument was built by CSL, IAS, MPS, MSSL/UCL, PMOD/WRC, ROB, LCF/IO with funding from the Belgian Federal Science Policy Office (BELSPO/PRODEX PEA 4000134088); the Centre National d’Etudes Spatiales (CNES); the UK Space Agency (UKSA); the Bundesministerium für Wirtschaft und Energie (BMWi) through the Deutsches Zentrum für Luftund Raumfahrt (DLR); and the Swiss Space Office (SSO).
        The development of SPICE has been funded by ESA member states and ESA. It was built and is operated by a multi-national consortium of research institutes supported by their respective funding agencies: STFC RAL (UKSA, hardware lead), IAS (CNES, operations lead), GSFC (NASA), MPS (DLR), PMOD/WRC (Swiss Space Office), SwRI (NASA), UiO (Norwegian Space Agency).
        The German contribution to SO/PHI is funded by the Bundesministerium für Wirtschaft und Technologie through Deutsches Zentrum für Luftund Raumfahrt e.V. (DLR), Grants No. 50 OT 1001/1201/1901 as well as 50 OT 0801/1003/1203/1703, and by the President of the Max Planck Society (MPG).
        The Spanish contribution is funded by AEI/MCIN/10.13039/501100011033/(RTI2018-096886-C5, PID2021-125325OB-C5, PCI2022-135009-2) and ERDF “A way of making Europe”; “Center of Excellence Severo Ochoa” awards to IAA-CSIC (SEV-2017-0709, CEX2021-001131-S); and a Ramón y Cajal fellowship awarded to DOS.
        The French contribution is funded by the Centre National d’Etudes Spatiales.
        Part of this work was supported by the Italian agreement ASI-INAF 2021-12-HH.0 “Missione Solar-C EUVST – Supporto scientifico di Fase B/C/D; Addendum N. 2021-12-HH.1-2024”. D.B. is funded under Solar Orbiter EUI Operations grant No. ST/X002012/1 and Hinode Ops Continuation 2022-25 grant No. ST/X002063/1. SML thanks Marco Stangalini, Gherardo Valori and Nawin Ngampoopun for useful discussion and suggestions that improved the manuscript. The authors acknowledge important suggestions from the anonymous referee who helped to improve the manuscript. This study has made use of SAO/NASA Astrophysics Data System’s bibliographic services. 
\end{acknowledgements}

%
%

\newpage
\begin{appendix}
\section{Alignments}\label{appendixalign}
The SPICE raster needs to be co-aligned with the HRI$_{EUV}$ sequence, since the pointing information in the SPICE L2 headers is not accurate. Using the \texttt{euispice\_coreg} python package (\citealt{dolliou2024}), we start building an HRI$_{EUV}$ synthetic raster by selecting from the HRI$_{EUV}$ time sequence the image closest in time to each SPICE exposure making the raster. Then, the HRI$_{EUV}$ image closest in time to each SPICE exposure is reprojected to the SPICE pixel positions along the slit. We then create the SPICE image in the \ion{Ne}{VIII} intensity by spectrally summing over the 25-pixel window. Finally, the SPICE image is co-aligned with the HRI$_{EUV}$ synthetic raster using a cross-correlation technique by using the SPICE \ion{Ne}{VIII} line as reference layer in the alignment process, as described e.g. in \citet{antolin2023}. The SPICE images for March 19, 2022 in the five spetrally-summed lines co-aligned with HRI$_{EUV}$ and ordered as function of temperature can be seen in Fig. \ref{lya_and_spice}.

To align the pointing of the HRI$_{Ly\alpha}$ image to the HRI$_{EUV}$ and SPICE images we cross-correlate it with the already-aligned SPICE Lyman$-\beta$ raster, following \citealt{auchere2020} (but see also \citealt{tian2009}). As a result, the HRI$_{Ly\alpha}$ image has been shifted of 19$\arcsec$ towards the West limb (Fig. \ref{lya_and_spice}).

The B$_{LOS}$ magnetogram also needs to be co-aligned to the HRI and SPICE images. \citet{auchere2020} suggest that the images taken by PHI/HRT can  be cross-correlated against SPICE raster images by exploiting the \ion{C}{I} continuum around 102 –
104 nm (range in which the SPICE Ly$\beta$ and \ion{O}{VI} lines fall), that is formed low enough in the atmosphere (above the temperature minimum, at a height of around 1000 km according to \citealt{vernazza1981}) that maps in this spectral emission should cross-correlate quite well with B$_{LOS}$ magnetograms (see Fig. 3 of \citealt{auchere2020}). To maximize the efficiency of the cross-correlation to align the PHI map (whose pixel scales are  0.5$\arcsec$), instead of using the SPICE rasters, which have spatial pixel scale of 4$\arcsec$ along the $X$ direction and 1.098$\arcsec$  along the $Y$ direction, we use as template layer the previously aligned HRI$_{Ly\alpha}$ image, which has instead pixel scale of 1.028$\arcsec$ along both spatial directions. The magnetic network is in fact apparent in both B$_{LOS}$ and HRI$_{Ly\alpha}$ images, and for this reason the HRI$_{Ly\alpha}$ has been used to align HRT B$_{LOS}$ maps by other authors (e.g. \citealt{nolke2023}; \citealt{kahil2020}). Therefore, the alignment chain applied to align the B$_{LOS}$ magnetogram is: HRI$_{EUV}$ to SPICE (through \ion{Ne}{VIII}) to HRI$_{Ly\alpha}$ (through Ly$\beta$) to HRT B$_{LOS}$. The final shift applied to the B$_{LOS}$ magnetogram is of 20$\arcsec$ towards the West limb and 72$\arcsec$ towards the South Pole (Fig. \ref{phi}).

\section{Intensity fluctuations}
We present here the intensity profiles for the vertical and horizontal cuts, V2, H1, and H2 (shown in Fig. \ref{qsfig}) from the HRI$_{EUV}$ dataset. The profiles display the dynamics of the DH fine structure over a $\sim$ 1 hr interval. 

\begin{figure}
    \centering
    \includegraphics[scale=.45,trim= 14 15 30 13,clip]{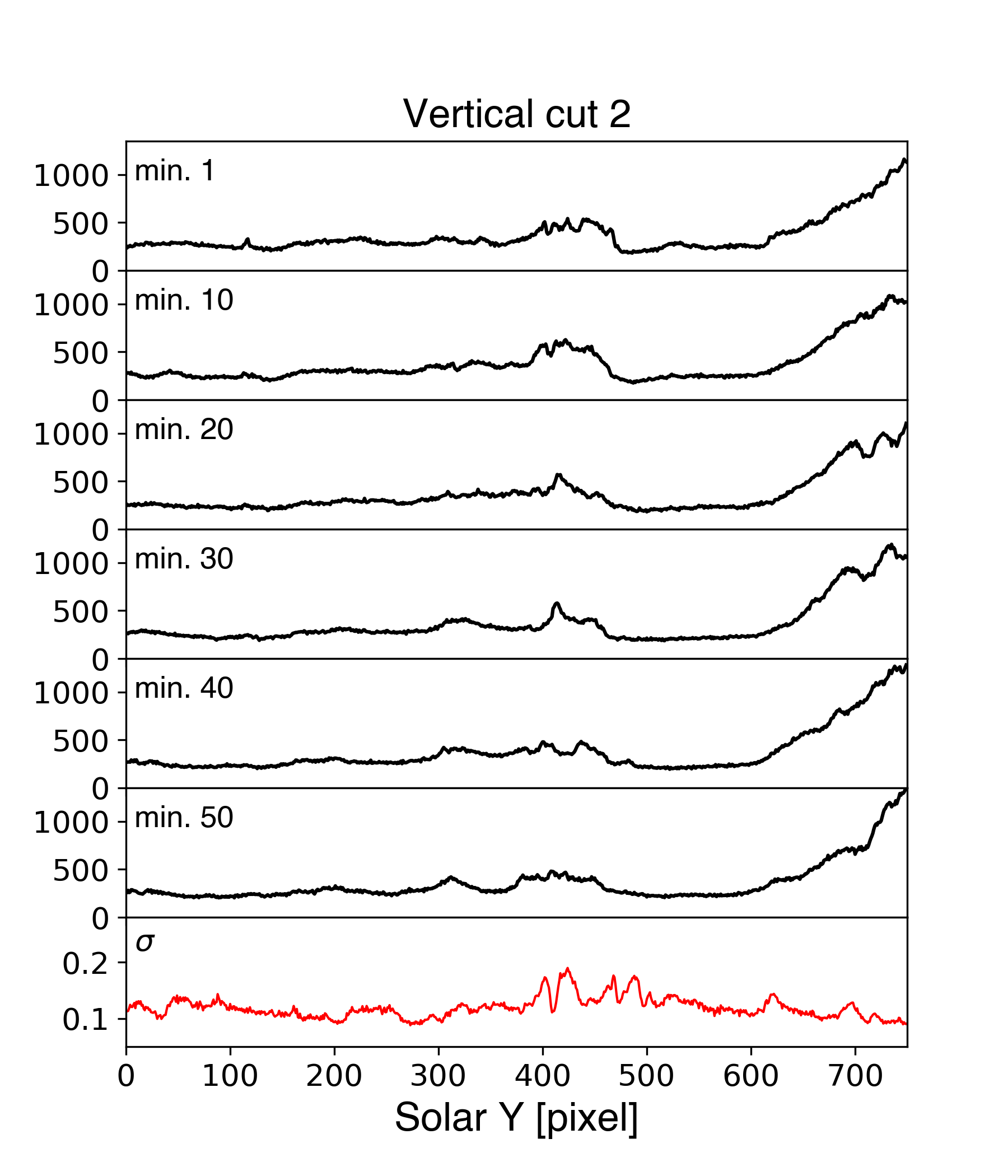}
    \includegraphics[scale=.45,trim= 14 15 30 13,clip]{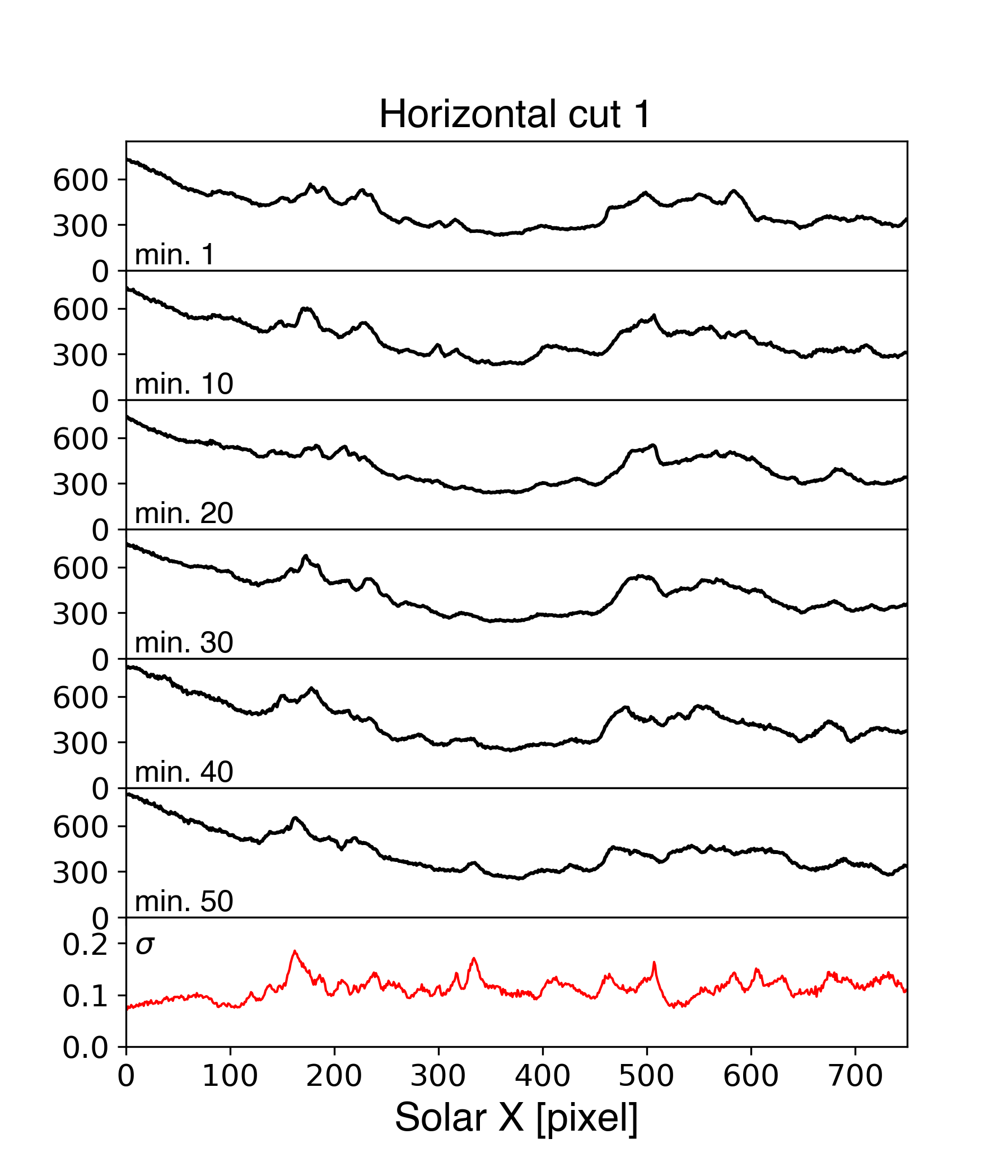}
    \includegraphics[scale=.45,trim= 14 15 30 13,clip]{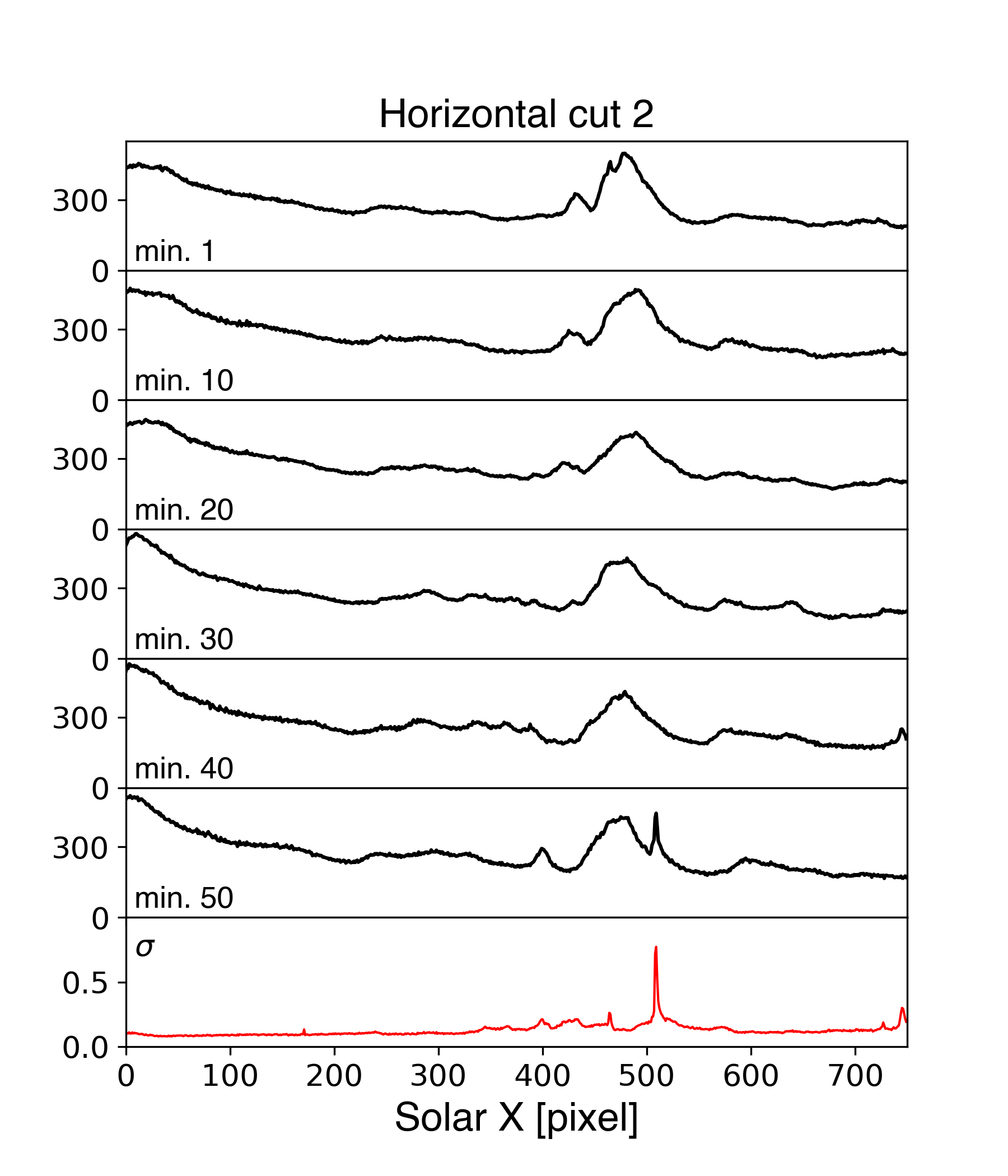}
    \caption{Intensity fluctuations (in DN/s) of the cuts V2, H1 and H2 shown in Figs. \ref{hrieuvfov} and \ref{qsfig}. The intensity profiles (black lines) are 1-min averages and are shown at a 10-min interval. The red lines are the standard deviations (normalized to the mean) of the slices for the 1-hour HRI$_{EUV}$ timeseries.} 
    \label{iprofiles2}
\end{figure}

\end{appendix}

\end{document}